# Localization property of a periodic chain of atoms with aperiodically coupled quantum dots

Atanu Nandy

Department of Physics, Acharya Prafulla Chandra College, New Barrackpore, Kolkata, West Bengal – 700 131, India

`atanunandy1989@gmail.com`

**Abstract.** The spectral landscape and the transport property of a translationally invariant network with side-coupled quantum dots are demonstrated within the tight-binding framework. For periodic environment band structure is demonstrated analytically in details. Moreover, if the side-coupling here follows a typical quasiperiodic Aubry-André-Harper type of modulation then such off-diagonal disorder invites an exotic spectral feature for this model quantum system. We perform an in-depth numerical analysis followed by the evaluation of the density of eigenstates and the inverse participation ratio. The description shows that this network creates a typical self-similar kind of multifractal pattern in the energy landscape. The impacts of the strength of such aperiodic connectivity and the slowness parameter are reported in this analysis. In the present era of advanced technology and lithography techniques all such non-trivial results definitely throw an achievable challenge to the experimentalists to study the localization of excitation in such network.

## 1      Introduction

The discovery of quantum Hall effect [1] has released a plethora of theoretical observations and experimental findings of different nontrivial topological phases [2] and this makes it to one of the most active research domains in the present era of condensed matter physics. Such systems offer nontrivial quantum phases, characterized by topological edge modes on their surface, which are insensitive against local perturbation. One of the most remarkable models in the context of topological phases was demonstrated by the pioneering work of Hofstadter [3]. His observation narrates a system of two-dimensional electron gas in a periodic lattice potential. When a strong magnetic perturbation is applied, the interplay of two length scales causes the energy bands to fragment into subbands creating a self-similar fractal pattern, known as the "Hofstadter butterfly." This central perception has made it feasible to realize the butterfly in graphene [4] and semiconductor superlattice [5] structures. Even reconfigurable quasiperiodic acoustic crystals have recently been reported to create Hofstadter butterfly and topological edge modes, taking the advancement beyond the realm of conventional electronic systems.

$AB_2$ kind of stub geometry which is essentially a periodic lattice with every alternate site attached to a side coupled quantum dot. This network is known to offer a central flat band along with two resonant subbands. In this article, we study the impact of aperiodic off-diagonal modulation on such periodic substrate. Initially, the Aubry-André-Harper (AAH) model was proposed as a paradigmatic model of an incommensurate one-dimensional system that exhibits a metal-insulator transition in parameter space. In this work we report that even such a minimal off-diagonal modulation can generate quantum butterfly pattern in the energy landscape that gives rise to plenty of rich physics.

## 2 Description of the model system

### 2.1 Tight-binding Hamiltonian

Let us refer to the Fig. 1 where a periodic $AB_2$ kind of stub geometry is cited. The standard tight-binding Hamiltonian for the demonstration of the system, written in the Wannier basis for the spinless non-interacting fermions, viz.,

$$H = \sum_n \varepsilon_n a_n^\dagger a_n + \sum_{<nm>} (t_{nm} a_n^\dagger a_m + h.c.)$$

The first term describes the potential information of the respective atomic site location and the off-diagonal term denotes the overlap integral between the nearest neighboring nodes. In our description, we take the on-site parameters of A and B types of sites as $\varepsilon_A$ and $\varepsilon_B$ respectively. Also, the hopping integral along the backbone is taken as t and that with the side-attached quantum dot is considered as λ.

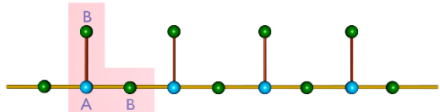

**Fig. 1.** A portion of an infinitely extended stub geometry. The shaded portion denotes the unit cell of the periodic lattice.

### 2.2 Band diagram of periodic lattice

The periodic environment of the $AB_2$ geometry leads to the expected delocalization of single particle eigenstates. The density of states (DOS) pattern can be evaluated using the real space decimation technique. A selected portion of the underlying lattice can be decimated out to transform it into an effective one-dimensional chain of identical atomic sites. The real space renormalization process within the tight-binding framework leads to a fixed point behavior of the parameters of the Hamiltonian from which one can easily obtain the electronic density of states with the aid of the stand-ard green's function formalism. The variation of the DOS with energy is plotted in the Fig. 2(a) where we notice two absolutely continuous subbands densely populated by resonant eigenfunctions along with the central spiky bound mode at E = 0. With the help of decimation technique, one can easily exploit the band dispersion relation for such periodic geometry. The energy momentum relation is plotted in the Fig. 2(b) where we see the two dispersive bands and one k-independent non-dispersive flat band at E = 0.

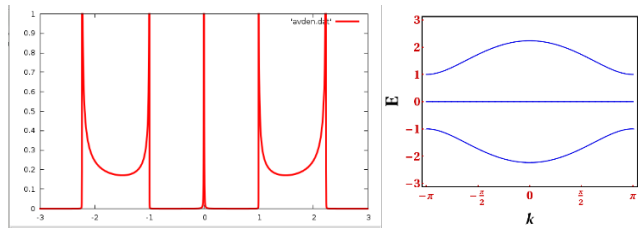

**Fig. 2.** (a) Variation of density of states with energy for the periodic stub lattice and (b) band dispersion of the same.

## 3 Impact of aperiodic side coupling

### 3.1 Aperiodic coupling scheme

To incorporate the discussion regarding the aperiodic side coupling we introduce an Aubry-André-Harper (AAH) modulation profile in the hopping integral λ that connects the quantum dots with the backbone. The aperiodic fashion of the overlap parameter for any n-th unit cell is governed by the following expression, viz.,

$$\lambda_n = \lambda_0[1 + \cos(\pi Q n^\beta a)]$$

Here $\lambda_0$ denotes the strength of the aperiodic modulation, a is the uniform lattice spacing and β denotes the slowness index. The parameter Q controls the frequency of this modulation.

### 3.2 Hofstadter butterfly

In this demonstration, we analyze the nature of the single particle energy spectrum of this geometry in presence of modulation in the side-coupling as a function of the modulation frequency for different strengths of the overlap integral. Interestingly however, very much similar with the AAH case, we see that the variation of Q creates self-similar kind of quantum butterfly in the energy landscape. The interplay between the modulation frequency and the strength of the coupling is a key aspect for the occurrence of butterfly spectrum. The pattern is full of energy gaps which is a characterizing feature of such aperiodic modulation. The in-gap states are robust within a window of moderate strength of the side-coupling. With the increase of strength of the intermediate gaps between the subbands gradually gets filled up by allowed states and the range of the energy spectrum has widen up. However, the butterfly is destroyed the moment we set fractional value of slowness index.

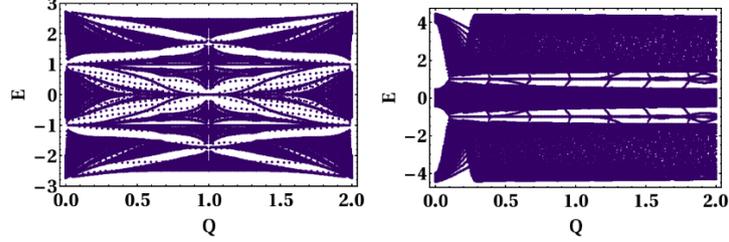

**Fig. 3.** (a) Self-similar Hofstadter butterfly pattern in the energy landscape and (b) the quantum butterfly pattern is destroyed for fractional slowness index.

### 3.3 Density of states and inverse participation ratio

The spectral canvas can be well demonstrated through the evaluation of density of states (DOS) profile as a function of energy of the injected projectile. The modulation in off-diagonal tunneling parameter invites a gateway to study the variation in the spectral landscape. The interesting competitive scenario between the uniform base hopping and the deterministic modulation in the side-coupling definitely offers an exotic spectral landscape.

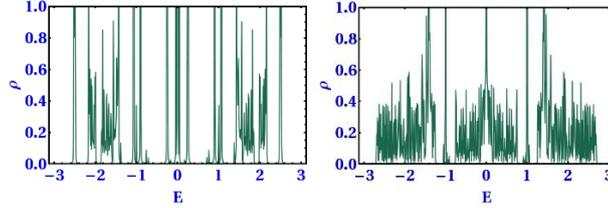

**Fig. 4.** (a) Variation of DOS with energy for unit slowness index and (b) the same for fractional slowness parameter.

For the sake of completeness of the discussion we have also evaluated the inverse participation ratio (IPR). To formulate the localization of a normalized eigenstate the inverse participation ratio is defined as

$$I = \sum_{i=1}^{L} |\psi_i|^4$$

It is known that for an extended mode IPR goes as $1/L$, but it approaches to unity for a localized state. The IPR plots are the proper justification of spectral profile.

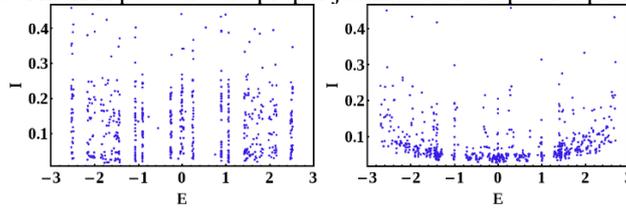

**Fig. 5.** (a) Variation of IPR with energy for unit slowness index and (b) the same for fractional slowness parameter.

## 4  Outlook

In conclusion, we have unraveled the spectral canvas of a periodic lattice with side coupled quantum dots. The side coupling can be designed in a suitable aperiodic fashion. The spectral competition between the axially periodic substrate and the off-diagonal modulation makes the spectral canvas interesting as well. Our analysis is corroborated by the numerical evaluation of density of states and the inverse participation ratio.